\documentclass{ws-mpla}

\begin{document}

\markboth{N F Naidu and M Govender} {Causal temperature profiles in
horizon-free collapse}

\catchline{}{}{}{}{}

\title{Causal temperature profiles in horizon-free collapse}

\author{\footnotesize N F Naidu\footnote{203507365@ukzn.ac.za}\,\, and M Govender\footnote{govenderm43@ukzn.ac.za} }

\address{Astrophysics and Cosmology Research Unit, School of Mathematics, University of KwaZulu-Natal, Durban, 4041,
South Africa. }

\maketitle

\pub{Received (Day Month Year)}{Revised (Day Month Year)}

\begin{abstract}
We  investigate the causal temperature profiles in a recent model of
a radiating star undergoing dissipative gravitational collapse
without the formation of an horizon. It is shown that this simple
exact model provides a physically reasonable behaviour for the
temperature profile within the framework of extended irreversible
thermodynamics.

\end{abstract}

\ccode{PACS Nos.: 04.20, 04.50+h.}

\section{Introduction}

The Cosmic Censorship Conjecture has continued to occupy center
stage within the realms of relativistic astrophysics. The final
outcome of the gravitational collapse of a star is still very much
open to debate with the discovery of models admitting naked
singularities.\cite{har1,har2} Various scenarios of gravitational
collapse have been considered in which the energy momentum tensor is
taken to be either a perfect fluid or an imperfect fluid with heat
flux and anisotropic pressure.\cite{3,nos,nolene} It is well known
that the collapse of reasonable matter distributions always lead to
the formation of a black hole in the absence of shear or in the case
of homogeneous densities. It has been shown that shearing effects
delay the formation of the apparent horizon by making the final
stages of collapse incoherent thus leading to the generation of
naked singularities.\cite{shear1} In this letter we revisit a
radiating stellar model proposed by Banerjee {\em et al}\cite{ban},
(hereafter referred to as the {\em BCD} model) in which the horizon
is never encountered. The interior matter distribution is that of an
imperfect fluid with heat flux and the exterior spacetime is
described by the radiating Vaidya metric.\cite{1} The junction
conditions required for the smooth matching of the interior and
exterior spacetimes across a four-dimensional time-like hypersurface
are solved exactly.

In this letter we investigate the physical viability of the {\em BCD
} model. In particular, we analyse the relaxational effects on the
temperature profiles within the framework of extended irreversible
thermodynamics. We are in a position to obtain exact solutions to
the causal heat transport equation for both the special case of
constant collision time as well as variable collision time. Our
results are in agreement with earlier thermodynamical investigations
of radiating stellar models. We find that relaxational effects
enhance the temperature at each interior point of the stellar
configuration. Our investigations show that the {\em BCD} model
displays physically reasonable temperature profiles throughout the
evolution of the star.

\section{The {\em BCD} radiating model revisited} In the {\em BCD}
model the following form of the metric for the interior spacetime is
assumed
\begin{equation} \label{1} ds^2 = -A^2(r,t) dt^2 + B^2(r,t)[dr^2 +
r^2 d \theta^2 + r^2 \sin^2 \theta d \phi^2]\,,
\end{equation} in which the metric functions $A$ and $B$ are yet to
be determined. The energy momentum tensor for the interior matter
distribution is given by \begin{equation} \label{2} T^{\mu \nu} =
(\rho + p) \, v^\mu v^ \nu + p g^{\mu \nu} + q^\mu v^\nu + q^\nu
v^\mu\,. \end{equation} The heat flow vector $q^\mu$ is orthogonal
to the velocity vector so that $q^\mu v_\mu = 0$. In order to
generate an exact model of radiative gravitational collapse, the
following ansatz was adopted for the metric functions in(\ref{1})
\begin{eqnarray}
A &=& a(r) \\
B &=& b(r)R(t) \end{eqnarray} which reduces the Einstein field
equations for the interior matter distribution to
\begin{eqnarray}
\rho &=& \frac{1}{R^2} \bigg[\frac{3}{a^2} {\dot R}^2 - \frac{1}{b^2} \left(\frac{2 b''}{b} - \frac{b'^{2}}{b^2} + \frac{4 b'}{r b} \right) \bigg] \label{e1}\\
p &=& \frac{1}{R^2} \bigg[-\frac{1}{a^2} (2 R {\ddot R} + {\dot R}^2
) + \frac{1}{b^2} \left(\frac{b'^2}{b^2} + \frac{2 a' b'}{a b} +
\frac{2}{r}
\left(\frac{a'}{a} + \frac{b'}{b} \right) \right) \bigg]\label{e2} \\
q^1 &=& - \frac{2 a' {\dot R}}{R^3 a^2 b^2}\,,
\label{e3}\end{eqnarray}where `dot' and `dash' indicate derivatives
with respect to time and the radial coordinate respectively. The
condition of pressure isotropy yields
\begin{equation} \label{cop}\frac{a''}{a} + \frac{b''}{b} -
2\frac{{b'}^2}{b^2} - 2\frac{a'b'}{ab} - \frac{a'}{ra} -
\frac{b'}{rb} = 0. \end{equation} Since the star is radiating energy
the exterior spacetime is described by the Vaidya metric given
explicitly in the form \begin{equation} \label{vaidya} d s^2 = -
\left(1 - \frac{2 M(v)}{\bar r} \right) d v^2 - 2 d \bar{r} d v +
\bar{r}^2 (d \theta^2 + \sin^2 \theta d\phi^2)\,, \end{equation}
where $v$ is the retarded time and $M(v)$ is the exterior Vaidya
mass. The junction conditions required for the smooth matching of
the interior metric (\ref{1}) and the exterior Vaidya metric
(\ref{vaidya}) across a time-like hypersurface ${\Sigma}$ are given
by
\begin{eqnarray}
(r B)_{\Sigma} &=& \bar{r}_{\Sigma} \\
p_{\Sigma} &=& (q^1 B)_{\Sigma} \label{p} \\
m_{\Sigma} &=& \Bigg[\frac{r^3 B {\dot B}^2}{2 A^2} - r^2 B' -
\frac{r^3 B'^2}{2 B} \Bigg]_{\Sigma}\,, \end{eqnarray} where
$m_{\Sigma}$ represents the total mass of the stellar configuration
of radius $r$ inside $\Sigma$. Utilising (\ref{e2}) and (\ref{e3})
in the boundary condition (\ref{p}) yields
\begin{equation} \label{bound} 2 R {\ddot R} + {\dot R}^2 + m {\dot
R} = n\,,
\end{equation} where $m$ and $n$ are constants. A simple particular
solution of (\ref{bound}) is \begin{equation} R(t) = - C t,
\end{equation} where $C
> 0$ is a constant of integration.
As pointed out in Ref.\refcite{ban} the mass-to-radius ratio,
${m_{\Sigma}}/ {\bar{r}_{\Sigma}}$, is independent of time. A
simple calculation yields\begin{equation} \frac{2
m_{\Sigma}}{\bar{r}_{\Sigma}} = \frac{2 m_{\Sigma}}{(r
B)_{\Sigma}} = 2 \bigg[\frac{C^2 r_0^2 b_0^2}{2 a_0^2} - \frac{r_0
b_0'}{b_0} - \frac{r_0^2 b_0'^2}{2b_0^2} \bigg]\,,
\label{mr}\end{equation} where $b(r_0) = b_0$ and $r_0$ defines
the boundary of the stellar configuration. It is interesting to
note that the parameters in (\ref{mr}) may be chosen so that $2
m_{\Sigma}/{\bar{r}_{\Sigma}} < 1$  in order to avoid the
appearance of horizon at the boundary.

\section{Causal temperature profiles} In this section we consider
the physical viability of the {\em BCD} model. In order to satisfy
the condition of pressure isotropy (\ref{cop}), the {\em BCD} model
assumes $b(r) = 1$ and
\begin{equation} A = a(r) = (1 + \xi_0 r^2) \end{equation} which in view of
(17-20) leads to  \begin{equation} m = - 4 \xi_0 r_0, \, \, n = 4
\xi_0 (1 + \xi_0 r_0^2), \,\, C = \frac{1}{2} \bigg(-|m| + (m^2 +
4 n)^{1/2} \bigg). \end{equation} The fluid volume collapse rate
is
\begin{equation} \Theta = \frac{3}{A}\frac{\dot B}{B} = \frac{3}{(1
+ \xi_0 r^2)t}\end{equation} which is the same in both the radial
and tangential directions in the absence of shear. The proper
stellar radius is given by
\begin{equation} r_p(t) = \int_0^{b_0}{Bdr} = -Ctb_0\,.\end{equation} Since the
star is collapsing we require that $C$ be positive which
corresponds to $-\infty < t < 0$. We further have \begin{equation}
C^2 < 4 \xi_0(1 + \xi_0 r_0^2) . \end{equation} The Einstein field
equations (\ref{e1})--(\ref{e3}) reduce to \begin{eqnarray}
\rho &=& \frac{3}{t^2 (1 + \xi_0 r^2)^2} \\
p &=& \frac{1}{t^2 (1 + \xi_0 r^2)^2} \bigg[\frac{4 \xi_0}{C^2} (1 + \xi_0 r^2) - 1 \bigg] \\
q^1 &=& - \frac{4 \xi_0 r}{(1 + \xi_0 r^2)^2} \frac{1}{C^2 t^3}\,.
\end{eqnarray} We note that all the above thermodynamical quantities
diverge as $t \rightarrow 0$. The regularity conditions $\rho
> 0, p
> 0$ and $\rho' < 0$, and $p' < 0$ together with the dominant energy condition, $(\rho - p) >
0$ and the more stringent requirement $(\rho + p) > 2|q|$ are all
satisfied when \begin{equation} \Bigg[1 - \frac{2 \xi_0 r}{C}
\Bigg]^2
> - \frac{2 \xi_0}{C^2} (1 - \xi_0 r^2).
\end{equation} \\
We can now write \begin{equation} 1 - \frac{2
m_{\Sigma}}{\bar{r}_{\Sigma}} = \Bigg[1 - \frac{C^2 r_0^2}{(1 +
\xi_0r_0^2)^2} \Bigg]\,. \end{equation} We note that that when
\begin{equation} \label{inf}C^2 < 1/r_0^2 + \xi_0^2 r_0^2 + 2
\xi_0\,,\end{equation} the boundary surface can never reach the
horizon. \cite{ban} Furthermore, the surface redshift is given by
\begin{equation} \label{red} 1 + z_{\Sigma} = \left(1 +
r_0\frac{b_0'}{b_0} + r_0{\dot b_0}\right)^{-1}\end{equation} which
diverges for an observer at infinity at the time of the appearance
of the horizon. For the {\em BCD} model (\ref{red}) reduces to
\begin{equation}1 + z_{\Sigma} = \left(1 - Cr_0\right)^{-1}
\end{equation} which diverges when $C = {1}/{r_0}$.
In order to avoid the divergence of the surface redshift we must
have \begin{equation} \label{C}{1}/{r_0^2} < C^2 < 1/r_0^2 +
\xi_0^2 r_0^2 + 2 \xi_0\,,\end{equation} where we have taken
(\ref{inf}) into account. The luminosity of the star as perceived
by an observer at infinity is given by
\begin{equation} \label{lum} L = -\frac{dm}{dv} = \frac{c^3r^3(1 +
\xi_0r^2 - rC)}{(1 + \xi_0r^2)^4}\end{equation} which is independent
of time. We now turn our attention to the evolution of the
temperature profiles of the {\em BCD} model. To this end we employ
the causal transport equation for the heat flux, which in the
absence of rotation and viscous stress is given by
\begin{equation}{\tau} h_a{}^b {\dot q}_{b} + q_a = -\kappa(D_a T +
T {\dot u}^a)\,, \label{cmc}\end{equation} where $\tau$ is the
relaxation time for the thermal signals. Setting $\tau = 0$ in the
above, we regain the so-called Eckart transport equations which
predict infinite propagation velocities for the dissipative fluxes.
For the line element (\ref{1}) the causal transport equation
(\ref{cmc}) reduces to
\begin{equation} \label{ca1} \tau(qB)_{,t} + A(qB) = -\kappa
\frac{(AT)_{,r}}{B}
\end{equation} which governs the behaviour of the temperature.
Setting $\tau = 0$ in (\ref{ca1}) we obtain the familiar Fourier
heat transport equation \begin{equation} \label{ca2} A(qB) = -\kappa
\frac{(AT)_{,r}}{B} \end{equation} which predicts reasonable
temperatures when the fluid is close to quasi--stationary
equilibrium. In order to study the evolution of the temperature in
the {\em BCD} model we employ the thermodynamic coefficients for
radiative transfer as outlined in Ref.\refcite{mg}. The thermal
conductivity takes the form
\begin{equation} \kappa =\gamma T^3{\tau}_{\rm c}\,,
\label{a28}\end{equation} where $\gamma$ ($\geq0$) is a constant
and ${\tau}_{\rm c}$ is the mean collision time between the
massless and massive particles. We further adopt the generalised
power--law behaviour for $\tau_{\rm c}$
\begin{equation} \label{a29} \tau_{\rm c}
=\left({\alpha\over\gamma}\right) T^{-\omega} \,,\end{equation}
where $\alpha$ ($\geq 0$) and $\omega$ ($\geq 0$) are constants.
The velocity of thermal dissipative signals is assumed to be
comparable to the adiabatic sound speed, which is satisfied if the
relaxation time is proportional to the collision time:
\begin{equation}\tau =\left({\beta\gamma \over \alpha}\right)
\tau_{\rm c}\,, \label{a30}\end{equation} where $\tau$ ($\geq
0$)is a constant. The constant $\beta$ is a measure of the
strength of relaxational effects, with $\beta=0$ giving the
noncausal case. Using the above definitions for $\tau$ and
$\kappa$, (\ref{ca1}) takes the form \begin{equation} \beta
(qB)_{,t} T^{-\omega} + A (q B) = - \alpha \frac{T^{3-\omega}
(AT)_{,r}}{B} \label{temp1} \,.\end{equation} The Eckart
temperature is readily obtained by setting $\beta = 0$ in
(\ref{temp1}). We are in a position to integrate (\ref{temp1}) for
the special case $\omega = 0$ which corresponds to constant
collision time and more interestingly, the case $\omega = 4$ which
gives a variable collision time. For constant collision time
($\omega = 0$), the causal temperature profile is given by
\begin{eqnarray} \label{no1} T^4(r,t) &=& \frac{8\beta
\xi_0\left[2(r_0^2 - r^2) + \xi_0(r_0^4 - r^4)\right]}{\alpha t^2
(1 + \xi_0r^2)^4} + \frac{8\xi_0\left[3(r^2 - r_0^2) + 3\xi_0(r^4
- r_0^4) + \xi_0^2(r^6 - r_0^6)\right]}{3\alpha t (1 +
\xi_0r^2)^4} \nonumber\\&&+
\left(\frac{L}{4\pi\delta}\right)\frac{1}{r_0^2c^2t^2}\left(\frac{1
+ \xi_0r_0^2}{1 + \xi_0r^2}\right)^4\,.\end{eqnarray} where $L$ is
given by (\ref{lum}) and $\delta$ is a constant. For $\omega = 4$
, the causal temperature is given by
\begin{eqnarray} \label{no2}
T^4(r,t) &=& \frac{8\beta\xi_0}{\alpha t^2(1 +
\xi_0r^2)^3}\left[\left(\frac{1 + \xi_0r_0^2}{1 + \xi_0r^2}\right)[8
+ \alpha t(1 + \xi_0r_0^2)]e^{\frac{8\xi_0}{\alpha t}\left(\frac{r^2
- r_0^2}{(1 + \xi_0r^2)(1 + \xi_0r_0^2)}\right)} - [8 + \alpha t(1 +
\xi_0r^2)]\right] \nonumber
\\ &&+ \frac{512\beta\xi_0e^{-\left(\frac{8}{\alpha t(1 +
\xi_0r^2)}\right)}}{\alpha^2t^3(1 + \xi_0r^2)^4}\left[{\mbox
Ei}\left(\frac{8}{\alpha t(1 + \xi_0r^2)}\right) - {\mbox
Ei}\left(\frac{8}{\alpha t(1 + \xi_0r_0^2)}\right)\right]\nonumber
\\  && + \left[\frac{1 + \xi_0r_0^2}{1 +
\xi_0r^2}\right]^4\frac{L}{(4\pi\delta)r_0^2c^2t^2}e^{\frac{8\xi_0}{\alpha
t}\left[\frac{r^2 - r_0^2}{(1 + \xi_0r^2)(1 +
\xi_0r_0^2)}\right]}\,,\end{eqnarray} where the exponential
integral ${\mbox Ei}(z)$ is defined as
\begin{equation} {\mbox Ei}(z) =
-\int_{-z}^{\infty}{\frac{e^{-t}}{t}dt}\,.\end{equation} We note
that the noncausal temperature $(\beta = 0)$ and causal
temperature are equal at the boundary $(r = r_0)$. Figure 1 shows
that the relaxational effects are dominant when the stellar fluid
is far from equilibrium (large values of $\beta$). In the case of
variable collision time, figure 2, we see that the causal
temperature is everywhere greater than the corresponding noncausal
temperature within the stellar interior. Furthermore, figures 1
and 2 indicate that the causal temperatures at late times (large
values of $\beta$) decrease more rapidly than the causal
temperatures when the star is close to quasi-static equilibrium.
This is in agreement with the perturbative results of
Ref.\refcite{her2} as well as the acceleration-free model studied
in Ref.\refcite{mart2}.
\begin{figure}[th]
\centerline{\psfig{file=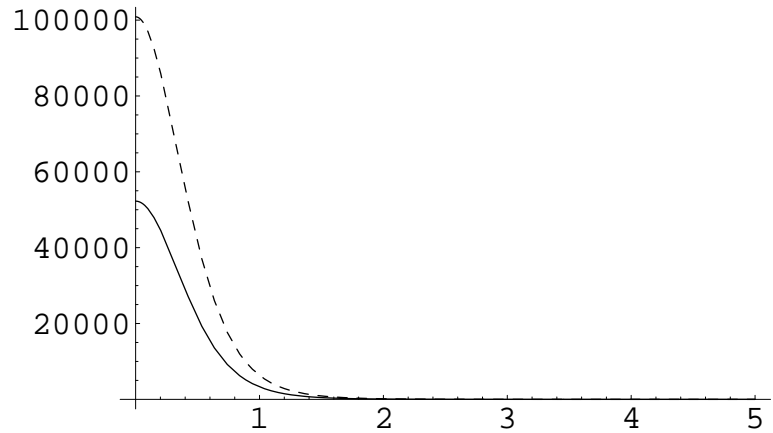,width=4.0in}} \vspace*{8pt}
\caption{Temperature profiles for constant collision time, (close to
equilibrium - solid line), (far from equilibrium - dashed line)
versus $r$.}
\end{figure}
\begin{figure}[th]
\centerline{\psfig{file=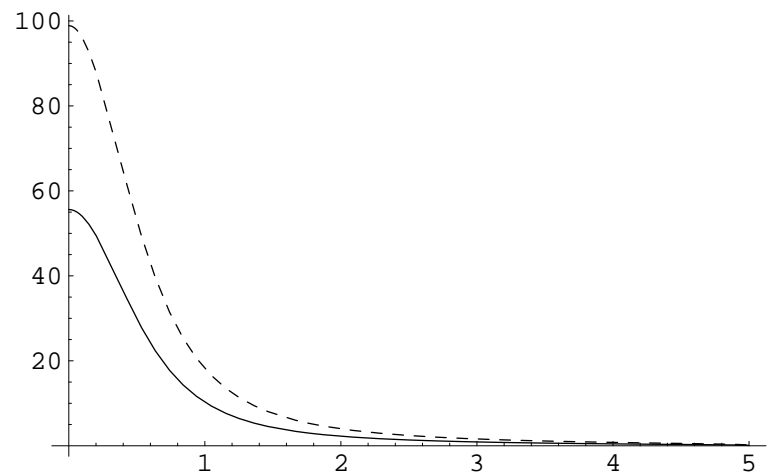,width=4.0in}} \vspace*{8pt}
\caption{Temperature profiles for variable collision time, (close to
equilibrium - solid line), (far from equilibrium - dashed line)
versus $r$.}
\end{figure}

\section{Concluding remarks}

We have investigated the physical viability of the {\em BCD} model
within the framework of extended irreversible thermodynamics. We
have shown that this simple model allows us greater insight into
the evolution of the temperature for different collision times.
More importantly, we were able to confirm earlier findings that
the causal temperature dominates the Eckart temperature within the
stellar core, even for variable collision time. As pointed out in
earlier treatments, the constant collision time approximation is
only valid for a limited period of the stellar
evolution.\cite{nolene} One expects that the collision time
between the particles making up the stellar fluid to change with
temperature. Such effects on the evolution of the temperature
profiles were clearly demonstrated with the variable collision
time solution. It must be pointed out that the truncation of the
transport equations leads naturally to an implicitly defined
temperature law. Such a temperature law may only be valid for a
limited period of collapse. What remains is to investigate the
behaviour of the temperature by employing the full transport
equation for the heat flux as well as to include the effects of
shear. The general framework for such an investigation has
recently been provided in Ref.\refcite{nos2} for a spherically
symmetric radiating star.


\begin{thebibliography}{0}

\bibitem{har1} T. Harada, H. Iguchi and K. Nakao, {\em Phys. Rev.}
{\bf D58}, 041502-1 (1998).

\bibitem{har2} H. Kudoh, T. Harada and H. Iguchi, {\em Phys. Rev.}
{\bf D62}, 104016-1 (2000).

\bibitem{3}  W. B. Bonnor, A. K. de Oliveira \& N.O. Santos, {\em Phys. Rep.} {\bf 181}, 269
(1989).

\bibitem{nos} L. Herrera and N. O. Santos, {\em Phys. Rep} {\bf
286}, 53 (1997)

\bibitem{nolene} N. F. Naidu, M. Govender and K. S. Govinder,
gr-qc/0509088 (2005).

\bibitem{shear1} P. S. Joshi, N. Dadhich and R. Maartens, {\em Phys.
Rev.} {\bf D65}, 101501 (2002).

\bibitem{ban} A. Banerjee, S. Chatterjee and N. Dadhich, {\em Mod.
Phys. Lett. A} {\bf 35}, 2335 (2002).

\bibitem{1}  P. C. Vaidya, {\em Proc. Ind. Acad. Sci} {\bf A33}, 264 (1951).

\bibitem{mg} S. D. Maharaj and M. Govender, {\em IJMP} {\bf D14}, 667 (2005).

\bibitem{her2} L. Herrera and N. O. Santos, {\em M.N.R.A.S} {\bf
287}, 161 (1997).

\bibitem{mart2} M. Govender, S. D. Maharaj and R. Maartens, {\em Class.
Quantum Grav.} {\bf 15}, 323 (1998).

\bibitem{nos2} L. Herrera and N. O. Santos, {\em Phys. Rev.}, {\bf
D70}, 084004 (2004).
\end{thebibliography}
\end{document}